\documentclass[twocolumn]{aastex631}

\usepackage[scale=1.08]{newtx}

\usepackage{url}
\usepackage{afterpage}
\usepackage{lipsum}
\usepackage{enumitem}
\usepackage{physics}
\usepackage{multirow}
\usepackage{textgreek}
\usepackage{booktabs}
\usepackage{xfrac}
\hypersetup{pdfnewwindow=true,
     colorlinks=true,
     linkcolor=blue, 
     citecolor=blue,
	 final=true}
	 
\usepackage{bm}
	 
\usepackage{etoolbox}
\makeatletter
\patchcmd{\NAT@citex}
  {\@citea\NAT@hyper@{%
     \NAT@nmfmt{\NAT@nm}%
     \hyper@natlinkbreak{\NAT@aysep\NAT@spacechar}{\@citeb\@extra@b@citeb}%
     \NAT@date}}
  {\@citea\NAT@nmfmt{\NAT@nm}%
   \NAT@aysep\NAT@spacechar\NAT@hyper@{\NAT@date}}{}{}
\patchcmd{\NAT@citex}
  {\@citea\NAT@hyper@{%
     \NAT@nmfmt{\NAT@nm}%
     \hyper@natlinkbreak{\NAT@spacechar\NAT@@open\if*#1*\else#1\NAT@spacechar\fi}%
       {\@citeb\@extra@b@citeb}%
     \NAT@date}}
  {\@citea\NAT@nmfmt{\NAT@nm}%
   \NAT@spacechar\NAT@@open\if*#1*\else#1\NAT@spacechar\fi\NAT@hyper@{\NAT@date}}
  {}{}
\makeatother

\newcommand{\JWST}{\textit{JWST}}

\newcommand{\hi}{H\,\textsc{i}}

\newcommand{\htwo}{H$_2$}

\newcommand{\oi}{O\,\textsc{i}}

\newcommand{\CI}{[C\,\textsc{i}]}
\newcommand{\CIA}{[C\,\textsc{i}]($2$--$1$)}

\newcommand{\COA}{CO($7$--$6$)}

\graphicspath{{figures/}}

\newcommand\target{A2744-45924}

\begin{document}

\title{\large Tentative detection of neutral gas in a Little Red Dot at $\bm{z=4.46}$}

\shortauthors{Akins et al.}
\shorttitle{}

\correspondingauthor{Hollis B. Akins} 
\email{hollis.akins@gmail.com}

\author[0000-0003-3596-8794]{Hollis B. Akins}
\altaffiliation{NSF Graduate Research Fellow}
\affiliation{Department of Astronomy, The University of Texas at Austin, Austin, TX 78712, USA}

\author[0000-0002-0930-6466]{Caitlin M. Casey}
\affiliation{Department of Physics, University of California, Santa Barbara, CA 93106, USA}
\affiliation{Cosmic Dawn Center (DAWN), Denmark}

\author[0000-0002-0302-2577]{John Chisholm}
\affiliation{Department of Astronomy, The University of Texas at Austin, Austin, TX 78712, USA}

\author[0000-0002-4153-053X]{Danielle A. Berg}
\affiliation{Department of Astronomy, The University of Texas at Austin, Austin, TX 78712, USA}

\author[0000-0003-3881-1397]{Olivia Cooper}
\altaffiliation{NSF Graduate Research Fellow}
\affiliation{Department of Astronomy, The University of Texas at Austin, Austin, TX 78712, USA}

\author[0000-0002-3560-8599]{Maximilien Franco} 
\affiliation{Universit\'{e} Paris-Saclay, Universit\'{e} Paris Cit\'{e}, CEA, CNRS, AIM, 91191 Gif-sur-Yvette, France}

\author[0000-0002-7530-8857]{Seiji Fujimoto}
\altaffiliation{Hubble Fellow}
\affiliation{Department of Astronomy, The University of Texas at Austin, Austin, TX 78712, USA}

\author[0000-0003-3216-7190]{Erini Lambrides}
\altaffiliation{NPP Fellow}
\affiliation{NASA Goddard Space Flight Center, 8800 Greenbelt Rd, Greenbelt, MD 20771, USA}

\author[0000-0002-7530-8857]{Arianna S. Long}
\affiliation{Department of Astronomy, University of Washington, Seattle, WA 98195, USA}

\author[0000-0002-6149-8178]{Jed McKinney}
\altaffiliation{Hubble Fellow}
\affiliation{Department of Astronomy, The University of Texas at Austin, Austin, TX 78712, USA}

\begin{abstract} 
\JWST\ has revealed a population of broad-line active galactic nuclei at $z>4$ with remarkably red colors, so-called ``Little Red Dots.'' 
Ubiquitous Balmer breaks suggest that they harbor old stellar populations in massive, compact host galaxies. 
We present ALMA observations of three LRDs at $z=3.10$, $4.46$, and $7.04$, targeting molecular and neutral gas via \COA\ and \CIA, respectively.
We do not detect CO in any target, placing conservative limits on the host molecular gas mass $\lesssim 1$--$5\times10^{10}$\,$M_\odot$. 
We report the tentative ($4.9\sigma$) detection of the \CIA\ line in \target\ ($z=4.46$), one of the brightest known LRDs. 
The \CI\ line is narrow (${\rm FWHM}\sim 80$\,km\,s$^{-1}$), implying a dynamical mass $\lesssim 10^{10}\,M_\odot$, adopting conservative limits for the galaxy size.  
The dynamical mass limit is significantly lower than expected from the local $M_{\rm BH}$--$M_{\rm dyn}$ relation, and is an order of magnitude below the stellar mass derived from SED fitting, potentially supporting a non-stellar origin of the Balmer break.
These results, while tentative, paint a picture of LRDs that is markedly different than typical high-$z$ quasars, which live in massive, gas-rich, and actively star-forming host galaxies. 
\end{abstract}

\section{Introduction}\label{sec:intro}

The formation of the first supermassive black holes (SMBHs), and their co-evolution with galaxies, remains a key question of modern astrophysics \citep{kormendyCoevolution2013}. 
The existence of luminous quasars as early as $z\gtrsim 7$ likely requires the existence of massive black hole seeds or accretion rates beyond the Eddington limit \citep{agarwalUbiquitous2012, pezzulliSuperEddington2016, inayoshiAssembly2020, fanQuasars2023}.
However, conducting a complete census of early SMBH growth is made complicated by the fact that active galactic nuclei (AGN) at $z>7$ are likely highly obscured \citep[e.g.][]{niQSO2020, gilliSupermassive2022}, limiting optical/near-IR observations. 
This is supported by quasar proximity zone measurements \citep{eilersDetecting2020, eilersDetecting2021, daviesEvidence2019} and the identification of several remarkably luminous obscured AGN at $z\gtrsim 7$ \citep{fujimotoDusty2022, endsleyRadio2022, endsleyALMA2022, lambridesUncovering2023} in relatively small survey areas.

A key result from \JWST\ is the discovery of an abundant population broad line AGN at $z>4$ \citep{harikaneJWST2023, taylorBroadLine2024}. 
Among these are the ``Little Red Dots'' (LRDs), first identified as broad-line H$\alpha$ emitters in NIRCam/WFSS data \citep{mattheeLittle2024}. 
The LRDs are characterized by a unique SED shape: blue in the rest-UV, but with a sharp turnover to a very red optical slope, suggesting significant dust attenuation \citep{labbeUNCOVER2023}. 
They appear contribute significantly to the overall AGN population, and bolometric luminosity function, at $z\sim 4$--$9$ \citep{greeneUNCOVER2024, kokorevCensus2024, kocevskiRise2024, akinsCOSMOSWeb2024}.  
While generally fainter than the UV-bright quasars, their black hole masses and bolometric luminosities can approach the regime of low-luminosity quasars ($L_{\rm bol} \gtrsim 10^{45}$\,erg\,s$^{-1}$), despite their apparently high abundance.

\begin{deluxetable*}{@{\extracolsep{7pt}}l@{}C@{}c@{}c@{}C@{}C@{}C@{}C@{}C@{}}[ht!]
\tabletypesize{\small}
\centering
\tablecaption{Summary of ALMA observations.}\label{tab:data}
\tablehead{\colhead{\multirow{2}{*}{Target}} & \colhead{\multirow{2}{*}{Redshift}}& \colhead{\multirow{2}{*}{Band}} & \colhead{Frequency Tuning} & \colhead{$T_{\rm int}$} & \colhead{Beam} & \colhead{RMS (cube)} & \colhead{RMS (cont)} \\[-0.7em] 
~ & ~ & ~ & \colhead{[GHz]} & \colhead{[min]} & \colhead{[$''\times ''$]} & \colhead{[$\mu$Jy/beam]} & \colhead{[$\mu$Jy/beam]}
}
\startdata
RUBIES-BLAGN-1 & 3.1034 & 5 & 194.1--197.7,~205.9--209.5 & 190.3 & 0.92\times0.68 & 102.5 & 9.7 \\
A2744-45924 & 4.4655 & 4 & 134.8--138.4,~146.9--150.5 & 98.5 & 1.56 \times 1.27 & 96.7 & 7.5 \\
A2744-QSO1 & 7.0367 & 3 & 87.6--91.2,~99.5--103.1 & 149.6 & 2.50 \times 1.85 & 68.3 & 5.3 \\
\enddata
\end{deluxetable*}
\vspace{-0.8cm}

The nature of the LRDs, however, remains unclear. 
In particular, they lack several key features common in AGN: they are generally weak in the mid-IR, indicating minimal hot dust emission from the torus \citep{williamsGalaxies2023, wangRUBIES2024, akinsCOSMOSWeb2024, liLittle2024}. 
They are also X-ray weak, with non-detections in deep X-ray stacks \citep{anannaXray2024,yueStacking2024,maiolinoJWST2024,lambridesCase2024}, and only two objects (out of hundreds) detected individually with \textit{Chandra} \citep{kocevskiRise2024}. 
Moreover, the host galaxies of LRDs are not well characterized; many LRDs exhibit Balmer breaks in their spectra \citep{wangRUBIES2024, labbeUnambiguous2024, furtakHigh2024}, which typically serves as a signature of an old stellar population. 
If the optical continuum in LRDs is dominated by an old stellar population, the implied stellar masses are large ($\gtrsim 10^{10\text{--}11}\,M_\odot$), and given their characteristic compactness, the required stellar mass \textit{densities} approach that of the densest nuclear star clusters \citep{baggenSizes2023, baggenSmall2024, akinsCOSMOSWeb2024}.
Curiously, the wavelength of the turnover from blue-to-red appears to occur consistently at the Balmer limit, $\sim 3800$\,\AA\ \citep{settonUNCOVER2024}, which is unexpected if the SED shape arises from a composite of galaxy+AGN components.

It has also been suggested that the Balmer break may arise not from old stars, but instead from absorption from nuclear gas with sufficient density to continuously populate the $n=2$ state via collisional excitation \citep{inayoshiExtremely2025}. 
In this scenario, the observed Balmer break is purely a product of the AGN structure. 
This is somewhat supported by the observations of velocity-offset Balmer absorption lines (e.g.~H\textalpha, H\textbeta) in the spectra \citep[e.g.][]{mattheeLittle2024, taylorBroadLine2024} , as well as the X-ray weak nature of the LRDs, which can originate from Compton-thick gas absorption \citep{maiolinoJWST2024}. 
However, current \JWST\ spectroscopy is not sufficiently deep to directly probe the spectrum around the break at high spectral resolution, limiting the possibility of any robust conclusions.

In this letter, we present ALMA spectroscopic observations of three LRDs at $z=3.10$--$7.04$, targeting the \COA\ and \CIA\ lines at 806.65 and 809.34 GHz. 
These far-infrared lines, tracing the molecular and atomic gas in the host galaxy interstellar medium (ISM), are routinely observed in high-$z$ quasars and dusty star-forming galaxies. 
While CO and \CI\ are typically fainter than [C\,\textsc{ii}]\,158\,$\mu$m, their similar frequencies make it possible to simultaneously trace multiple gas phases to better characterize the host galaxies of LRDs. 

\defcitealias{wangRUBIES2024a}{W24b}
\defcitealias{labbeUnambiguous2024}{L24}
\defcitealias{furtakHigh2024}{F24}

\section{ALMA observations and data reduction}

We targeted three LRDs with existing (spectroscopic) redshifts from \JWST: RUBIES-BLAGN-1 at $z=3.1034$ \citep[][hereafter \citetalias{wangRUBIES2024a}]{wangRUBIES2024a}, A2744-45924 at $z=4.4655$ \citep[][hereafter \citetalias{labbeUnambiguous2024}]{labbeUnambiguous2024}, and A2744-QSO1 at $z=7.0367$ \citep[][hereafter \citetalias{furtakHigh2024}]{furtakHigh2024}. 
These objects were selected for their relative brightness and accessibility to ALMA. 
RUBIES-BLAGN-1 and A2744-45924 are among the brightest LRDs known, and A2744-QSO1, while intrinsically fainter and at much higher redshift, is strongly lensed, with images A ($\mu=6.2$) and B ($\mu=7.3$) fitting into the same ALMA pointing. 

\begin{figure*}
	\includegraphics[width=\linewidth]{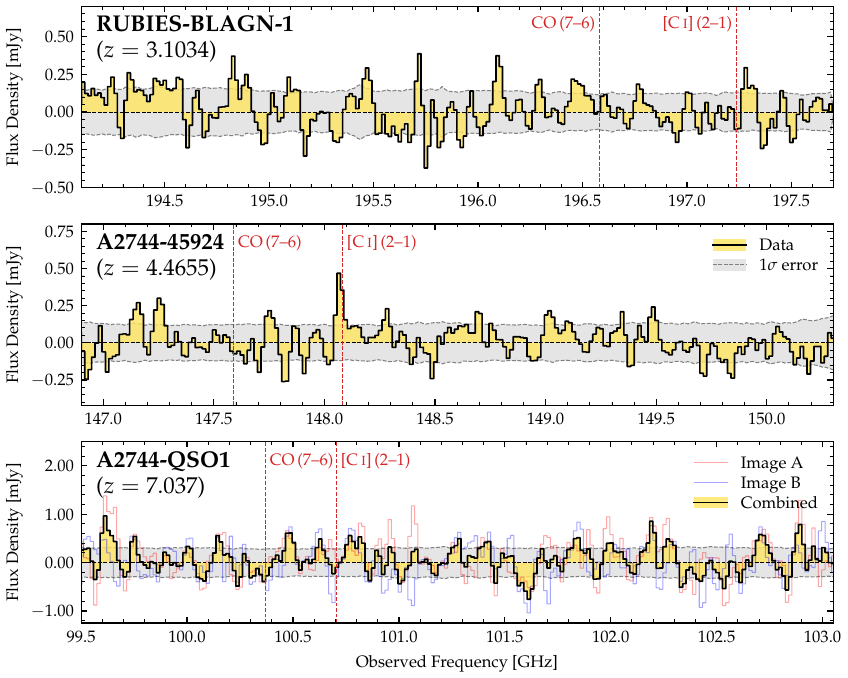}
	\caption{Extracted spectra covering \COA\ and \CIA\ for the three LRD targets. Spectra are extracted from $r=0\farcs5$ apertures and corrected to account for the total beam. The grey shaded region shows the $1\sigma$ error, measured from random apertures in each channel of the cube. For A2744-QSO1, we show the extracted spectra at the positions of image A and B, as well as the co-added spectrum. \COA\ is not detected in any of these three LRDs; however, we identify a tentative feature at the wavelength of \CIA\ for \target.}\label{fig:spectra}
\end{figure*}

The ALMA observations (PID 2024.1.01085.S) were tuned to detect \COA\ and \CIA\ at rest-frame frequencies of 806.65 and 809.34 GHz, respectively. 
Integration times were determined by scaling the measured black hole masses to host galaxy bulge masses assuming local relations \citep{kormendyCoevolution2013}, converting to CO luminosities, and scaling down by a conservative factor. 
The Common Astronomy Software Applications package \citep[CASA;][]{mcmullinCASA2007} was used for reduction, calibration, and imaging. 
We adopt the standard ALMA pipeline results and perform custom imaging using the \texttt{tclean} task in CASA. 
We use the \texttt{auto-multithresh} automasking algorithm \citep{kepleyAutomultithresh2020} using the standard recommended parameters.\footnote{As detailed in Section 9.39 of the ALMA pipeline User's Guide: \url{https://almascience.nrao.edu/documents-and-tools/cycle10/alma_pipeline_users_guide_2023}}
We also explored manually masking in a circular aperture around the known positions of each target, and find consistent results using this method. 
All images are made using natural weighting to maximize sensitivity, and datacubes are produced at the native channel width of each observation.
All maps are primary beam corrected.

Table~\ref{tab:data} summarizes the ALMA observations of the three LRDs. 
Between the three observations in bands 3, 4, and 5, the beam size ranges from $\sim 0\farcs9$--$2\farcs5$, and the RMS ranges from 68--103 $\mu$Jy/beam (measured in the primary beam \textit{uncorrected} datacubes).

\section{Results}\label{sec:results}

\subsection{CO(7--6) and [C\,\textsl{\textsc{i}}](2--1) line spectra}

Figure~\ref{fig:spectra} shows the extracted 1D spectra for the three targets. 
We mark the expected positions of the \COA\ and \CIA\ lines in red, based on the \JWST-derived redshifts.
We note that we adopt the redshifts derived from NIRSpec medium or high-resolution spectroscopy ($R\sim 1000$--$2700$) or NIRCam WFSS ($R\sim 1600$). 
These redshifts should be accurate to $\lesssim 30$ km\,s$^{-1}$; this is not necessarily true for redshifts derived from the PRISM, which can have a systematic offset up to $\sim 300$\,km\,s$^{-1}$ \citep{deugenioJADES2025}, making ALMA line identification difficult. 
For the multiply-imaged object A2744-QSO1, we show the extracted spectra at the positions of images A and B, and the coadded spectrum. 

No lines are detected at the expected wavelength of \COA\ in the three targets. 
We identify a tentative feature at the position of \CIA\ in \target, though we do not see a similar feature in RUBIES-BLAGN-1 or A2744-QSO1.

\begin{figure*}
\includegraphics{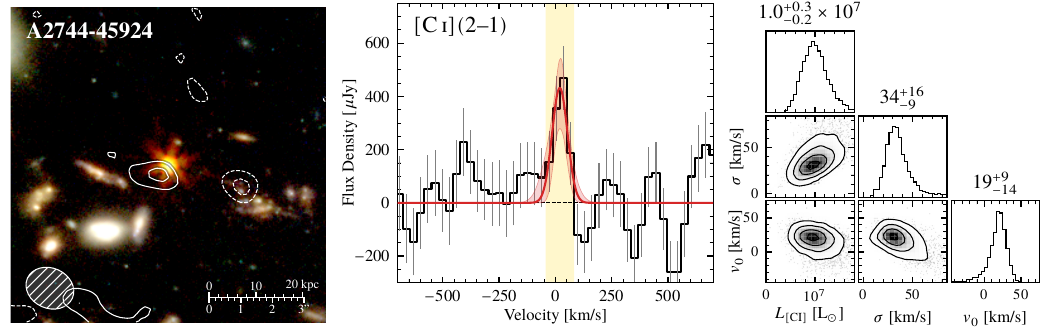}
\caption{The tentative \CIA\ detection in \target. \textbf{Left:} 10'' $\times$ 10'' \JWST/NIRCam RGB cutout around \target\ from the UNCOVER survey \citep{bezansonJWST2022}. We overlay the \CI\ moment-0 map as white contours ($-3$, $-2$, $2$, and $3\sigma$). The beam size ($1\farcs56\times1\farcs27$) is indicated by the ellipse in the bottom left. \textbf{Center:} The extracted 1D spectrum of \CIA\ at the peak position in the moment-0 map. Error bars show the 1$\sigma$ uncertainty derived from random aperture measurements across each channel in the cube. The red line and shaded region shows our best-fit Gaussian line profile fit with \texttt{emcee}. \textbf{Right:} Corner plot of the posterior line properties from our MCMC fit. }\label{fig:CI}	
\end{figure*}

\subsection{Dust continuum}

We do not detect any target in the continuum observations, consistent with the overall LRD population, which are undetected at far-infrared wavelengths \citep{labbeUNCOVER2023, akinsCOSMOSWeb2024}.  
We derive upper limits on the dust mass in the optically-thin, Rayleigh-Jeans tail of the SED, following the methodology of \citet{scovilleISM2016}.
Assuming a mass-weighted dust temperature $T_{\rm dust} = 25$ K and dust emissivity index $\beta=1.8$, we estimate dust mass limits of $M_{\rm dust} < 7\times 10^7$, $<1.4\times 10^8$,  and $<6\times 10^7\,M_\odot$ for RUBIES-BLAGN-1, A2744-45924, and A2744-QSO1, respectively.

\subsection{Detection of [C\,\textsl{\textsc{i}}](2--1) in \target}

We identify a $\sim 3.9\sigma$ peak in the spectrum at the expected frequency of \CIA\ in \target. 
Figure~\ref{fig:CI} shows the moment-0 map integrated over the four channels at this peak and the 1D spectrum extracted at the peak position.
The integrated signal-to-noise over these channels is $4.9\sigma$.

The peak position of the \CIA\ detection is located $\sim 0\farcs4$ to the SE of \target. 
Given the slight positional offset, we consider the possibility that the detection is spurious. 
Following \citet{ivisonSCUBA2007}, we estimate the positional uncertainty of the ALMA observations from the beam size $\theta$ and signal-to-noise ratio. 
Random positional offsets of distance $r$ will occur following the probability distribution $P(r) = r\,\exp(-r^2/(2\sigma^2))$, where $\sigma$ is the typical offset in RA/Dec, given by $\sigma = 0.6\,\theta\,(S/N)^{-1} \approx 0\farcs25$. 
Following this probability distribution, the expectation value of $r$ is $\approx 0\farcs32$, and offsets greater than $0\farcs4$ will occur $\sim 30\%$ of the time. 
We therefore consider the positional offset to be negligible, statistically consistent with emission from phase center.

We fit the spectrum with a Gaussian line profile using the \texttt{emcee} Python package. 
The right panel of Figure~\ref{fig:CI} shows the resulting posterior estimates of the line luminosity, width, and velocity offset from the nominal redshift of $z=4.4655$. 
We measure a total line flux of $I_{[{\rm C}\,\textsc{i}]} = 36.9_{-8.4}^{+10.6}$\,mJy\,km\,s$^{-1}$ and a luminosity of $L_{[{\rm C}\,\textsc{i}]} = 1.0^{+0.3}_{-0.2}\times 10^7\,L_\odot$. 
The line center is offset $19^{+9}_{-14}$\,km\,s$^{-1}$ from the nominal redshift, consistent with 0, and we derive a line FWHM of $80^{+38}_{-22}$\,km\,s$^{-1}$. 
The line width is remarkably narrow, and in fact is unresolved; the velocity resolution for the ALMA observations is $\sim 70$ km\,s$^{-1}$.
The narrow line width implies a remarkably low dynamical mass, which we return to in \S\ref{sec:dynamics}.

\subsection{Molecular gas mass}

We estimate the molecular gas masses (or upper limits) of the three LRDs using the \COA\ and \CI\ lines. 
Molecular gas masses are commonly derived from CO observations, though the use of high-$J$ CO lines imparts additional uncertainty from the CO excitation ladder. 
The \CI\ lines have proven effective tracers of the total molecular gas when low-$J$ CO is inaccessible \citep{carilliCool2013}. 
This is thought to be due to the stratified chemical structure in photodissociation regions (PDRs), in which C$^0$ occupies a very narrow transition layer between C$^+$ and CO \citep{wolfirePhotodissociation2022}. 
We perform both calculations in order to place conservative limits on the H$_2$ mass in the LRDs.

For CO, we provide two estimates of the upper limit on the H$_2$ mass. 
First, we adopt a standard set of assumptions for quasar host galaxies: a CO-to-H$_2$ mass conversion factor $\alpha_{\rm CO} = 0.8\,M_\odot\,({\rm K}\,{\rm km}\,{\rm s}^{-1}\,{\rm pc}^2)^{-1}$ and a relatively steep CO SLED, such that $L'_{{\rm CO}(7-6)}/L'_{{\rm CO}(1-0)} = 0.60$, similar to \citet{venemansMolecular2017}.
We assume CO line widths of 300 km\,s$^{-1}$ to translate the upper limits in terms of flux density to luminosity. 
These assumptions give us 3$\sigma$ molecular gas mass limits of $M_{{\rm H}_2,{\rm CO}} <4\times10^8$, $<9\times10^8$, and $<2\times10^9\,M_\odot$ for RUBIES-BLAGN-1, A2744-45924, and A2744-QSO1, respectively. 
Adopting a more conservative set of assumptions for CO-to-H$_2$ conversion factor of $\alpha_{\rm CO} = 4.5\,M_\odot\,({\rm K}\,{\rm km}\,{\rm s}^{-1}\,{\rm pc}^2)^{-1}$, the canonical value for the Milky Way, and a shallower CO SLED such that $L'_{{\rm CO}(7-6)}/L'_{{\rm CO}(1-0)} = 0.1$, approximately the lower bound observed for SMGs \citep*{caseyDusty2014}, yields larger molecular gas mass limits by a factor of $\sim 30$. 
Specifically, we derive $M_{{\rm H}_2,{\rm CO}} <10^{10}$, $<3\times10^{10}$, and $<5\times10^{10}\,M_\odot$ for RUBIES-BLAGN-1, A2744-45924, and A2744-QSO1, respectively.

For \CI, we adopt a \CI-to-H$_2$ conversion factor of $\alpha_{[{\rm C}\,\textsc{i}](2-1)} = 34$ from \citet{crockerI102019}. 
For \target, we derive a \CI\ luminosity of $L'_{[{\rm C}\,\textsc{i}]} = 5.8^{+1.7}_{-1.4}\times10^8$\,K\,km\,s$^{-1}$\,pc$^2$, which corresponds to a molecular gas mass of $M_{{\rm H}_2} = 2.0^{+3.1}_{-1.2}\times 10^{10}\,M_\odot$. 
We note that this accounts for uncertainty in $\alpha_{[{\rm C}\,\textsc{i}]}$ at the level of 0.4 dex. 
For RUBIES-BLAGN-1 and A2744-QSO1, we derive 3$\sigma$ limits of $M_{{\rm H}_2,[{\rm C}\,\textsc{i}]} < 10^{10}$ and $<4\times10^{10}\,M_\odot$. 
These limits are similar to the conservative limits from \COA\, but depend only on the empirically-derived $\alpha_{[{\rm C}\,\textsc{i}](2-1)}$ conversion factor, and not on the (highly uncertain) CO line SLED.

\subsection{Dynamical mass}\label{sec:dynamics}

We estimate the dynamical mass in \target\ based on the \CI\ line kinematics. 
The dynamical mass is estimated as 
\begin{equation}
	M_{\rm dyn} = \frac{5 R_e \sigma^2}{G}
\end{equation}
where $G$ is the gravitational constant, $R_e$ is the effective radius, and $\sigma$ is the velocity dispersion of the gas, in this case from the \CI\ line, $\sigma_{\rm gas} = 34^{+16}_{-9}$\,km\,s$^{-1}$. 
However, the relevant radius is not obvious, as the \CI\ line is spatially unresolved. 
\citet{labbeUnambiguous2024} fit 2D Sersic models to the \JWST/NIRCam imaging and estimate the effective radius $R_e\lesssim 70$ pc.
However, if the \CI\ originates from larger scales than the NIRCam emission (which may be predominantly from the AGN), this size would be too small. 
\citet{labbeUnambiguous2024} also find evidence for a faint, extended ($R_e \sim 700$ pc) structure alongside the dominant unresolved component. 
We adopt this size for our fiducial measurement of the dynamical mass, for which we derive $M_{\rm dyn} = 1.4^{+1.7}_{-0.6}\times 10^9\,M_\odot$. 
We note that this accounts for the unknown inclination angle via a factor of 3/2, which is the reciprocal of the expectation value of $\sin^2 i$. 

Given the unknown size of the \CI-emitting gas, we also conservatively quote an upper limit on the dynamical mass assuming a size of $R_e < 4.8$ kpc, which corresponds to the half-width at half-maximum of the ALMA beam.
This yields a dynamical mass limit of $M_{\rm dyn} < 9.4^{+11.4}_{-4.9}\times 10^9\,M_\odot$.  
The dynamical mass limit, even adopting a conservative limit on the size, is at least an order of magnitude below the stellar mass for \target\ derived by \citet{labbeUnambiguous2024} from SED fitting to the Balmer break, which we discuss further in Section~\ref{sec:discussion}.

\begin{figure}
\includegraphics[width=\linewidth]{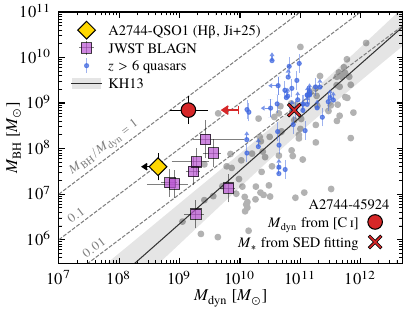}
\caption{Black hole mass vs.~dynamical mass \JWST\ AGN at $z>4$, including \target, compared to local SMBHs and $z>6$ quasars. We plot local SMBHs and the best-fit relation from \citet{kormendyCoevolution2013} in grey. We show $z>6$ quasars in blue, with ALMA [C\,\textsc{ii}] dynamical mass measurements \citep{izumiSubaru2019}. We also show other \JWST\ BLAGN with dynamical masses measured from high-resolution spectroscopy \citep{maiolinoJADES2023, ublerGANIFS2023}, including A2744-QSO1 \citep{jiBlackTHUNDER2025}. Our dynamical mass measurement for \target\ is comparable to the black hole mass, and significantly above the local relation.}\label{fig:MBH_Mdyn}
\end{figure}

\section{Discussion}\label{sec:discussion}

\subsection{Implications for the host galaxies of LRDs} 

The limits we place on the host galaxy dynamical and molecular gas masses allow us to better place the three targeted LRDs in context among high-$z$ galaxies and quasars. 
It has been suggested that the LRDs represent the obscured phase of quasar evolution \citep{schindlerBroadline2024, aritaNature2025} or are actually massive, compact galaxies \citep{baggenSmall2024}. 
In fact, the large black hole masses reported for two of our targets ($M_{\rm BH} \sim 1$--$4\times10^8\,M_\odot$ and $\sim 10^9\,M_\odot$ for RUBIES-BLAGN-1 and \target, respectively) already suggests some similarity to quasars, and the strong Balmer break in both objects indicates the presence of a massive stellar population. 
However, the CO and \CI\ observations imply low gas fractions: adopting typical assumptions for quasar host galaxies, our derived gas mass limits imply $f_{\rm gas} = M_{\rm gas} / (M_\star + M_{\rm gas}) < 20\%$ for RUBIES-BLAGN-1 and $<1\%$ for \target. 
This is in sharp contrast to high-$z$ quasars \citep{venemansMolecular2017, decarliALMA2018, izumiSubaru2019} and luminous obscured AGN \citep{fujimotoDusty2022, endsleyALMA2022, lambridesUncovering2023} which live in massive, gas-rich, and actively star-forming host galaxies.

The dynamical mass estimate for \target\ is remarkably low, comparable to the black hole mass derived from the broad H$\alpha$ line \citepalias{labbeUnambiguous2024}. 
This suggests that the SMBH in \target\ is overmassive relative to the host, even in terms of the dynamical mass. 
Figure~\ref{fig:MBH_Mdyn} shows black hole mass vs.~dynamical mass for \target\ alongside local SMBHs \citep{kormendyCoevolution2013}, high-$z$ quasars \citep{izumiSubaru2021} and recent JWST-discovered AGN \citep{maiolinoJADES2023, ublerGANIFS2023}. 
We also show A2744-QSO1; though not detected in \CI\ or CO, the dynamical mass has recently been measured from the narrow component of H$\beta$ via high-resolution \JWST\ IFU spectroscopy \citep{jiBlackTHUNDER2025}. 
They derive a black hole-to-dynamical mass ratio of $\gtrsim 0.1$, similar to this work. 
Our dynamical mass measurement positions \target\ significantly above the local $M_{\rm BH}$--$M_{\rm dyn}$ relation. 
The existence of such overmassive black holes may be a signature of high-mass seeds \citep[e.g.][]{agarwalUbiquitous2012} or super-Eddington growth \citep[e.g.][]{huSupercritical2022}.

Moreover, our derived dynamical mass for \target\ is more than an order of magnitude below the stellar mass derived in \citetalias{labbeUnambiguous2024} ($M_\star \sim 8\times 10^{10}\,M_\odot$), which is based on the assumption that the strong Balmer break observed in the NIRSpec spectrum is associated with an old stellar population. 
Provided the \CI\ detection traces the overall kinematics of the host galaxy, our results suggest that that the Balmer break in \target\ \textit{cannot} originate from stellar emission. 
Instead, it seems plausible that the Balmer break originates from extremely dense gas close to the AGN \citep[e.g.][]{inayoshiExtremely2025}. 
This is somewhat corroborated by the detection of velocity-offset H$\alpha$ absorption lines, as well as remarkably strong \oi\,$\lambda$8446 emission in \target\ \citepalias{labbeUnambiguous2024}. 
The latter can arise from resonance flourescence of Ly$\beta$ in extremely dense gas, due to the similar energy levels between \oi\ and \hi\ \citep[see further discussion in][]{inayoshiExtremely2025}.

\subsection{Origin of the [C\,\textsc{i}] emission}

The detection \CI\ in \target, without a corresponding detection of \COA, is distinct from high-$z$ quasars \citep[e.g.][]{venemansMolecular2017}, dusty star-forming galaxies \citep[e.g.][]{yangMolecular2017}, and even main-sequence galaxies at lower-redshift \citep[e.g.][]{valentinoCO2020}. 
Moreover, the molecular gas mass as estimated from the \CI\ line is more than an order of magnitude larger than the dynamical mass measured from the same line, perhaps suggesting that the mechanism exciting \CI\ in \target\ may be different than in the local galaxies used to calibrate the \CI-to-H$_2$ relation \citep[e.g., PDRs;][]{papadopoulosCI2004}. 
Such conditions have been observed in the circumnuclear disk of the local radio galaxy Centaurus A: \citet{israelMolecular2014} measure \CIA/\COA\ $\gtrsim 6$, a ratio larger than any known AGN or starburst galaxy and inconsistent with their PDR modeling.

There are a number of possible scenarios that could produce elevated \CI\ relative to CO, particularly at high redshift, where extreme ISM conditions occur more frequently \citep[e.g.][]{cameronJADES2023, katzFirst2023}.
The interpretation of this result depends in part on whether the \CI\ indeed traces \htwo\ gas, as is thought to be the case in PDRs, or primarily \hi\ gas. 
In the former scenario, the \htwo\ gas must be ``CO-dark,'' as in translucent clouds \citep{magnaniVariation1998, bolattoCOtoH22013}. 
One possibility is a super-solar C/O abundance, which would leave more of the carbon in its atomic phase, unable to form into CO, therefore increasing the C/CO column density ratio.
A super-solar C/O has been tentatively reported in another LRD at $z=7$ \citep{akinsStrong2025}, though no robust abundance measurements have been made.    
At the same time, the C/CO column density ratio can be increased by dissociation of the CO molecules. 
CO dissociation can occur in shocks, driven by outflows from the AGN \citep[e.g.][]{saitoKiloparsecscale2022}, or in X-ray dominated regions \citep[XDRs;][]{meijerinkDiagnostics2005, meijerinkDiagnostics2007, wolfirePhotodissociation2022}. 
However, we note that \target\ is undetected in deep (2.14 Ms) \textit{Chandra} X-ray observations (PID 23700107; PI: A.~Bogdan), consistent with the majority of LRDs and \JWST-selected AGN at $z>4$, \citep[e.g.][]{anannaXray2024, maiolinoJWST2024, lambridesCase2024}.
If the strong \CI\ is indeed driven by XDRs, it would have to be in regions of very high density, sufficient to absorb the X-rays as well.

It is also possible that the \CI\ primarily traces \hi\ gas rather than \htwo. 
Indeed, the non-stellar Balmer break interpretation \textit{requires} abundant \hi\ gas, at very high density, sufficient to continuously populate the $n=2$ state via collisional excitation \citep{inayoshiExtremely2025}. 
At these extreme densities ($n_H > 10^9$\,cm$^{-3}$) and temperatures ($T>8000$\,K), \htwo\ is dissociated via collisions with \hi\ gas. 
However, the critical density of the \CIA\ line is $<1000$\,cm$^{-3}$, far below the density required to produce the Balmer absorption. 
The \CI\ would therefore, in this scenario, arise from a lower-density component of the \hi\ gas responsible for producing the observed Balmer break and H$\alpha$ absorption lines in \target.
The formation of these lower density atomic clouds, without a significant molecular component, may be aided by strong FUV Lyman-Werner radiation fields, which can dissociate \htwo\ \citep{stecherPhotodestruction1967}; this argument is often invoked to prevent cooling and fragmentation in models of primordial gas clouds forming into ``direct collapse black holes'' \citep[e.g.][]{brommFormation2003}. 
X-ray radiation can have a similar effect by heating gas \citep{jeonRadiative2014}, but it is less clear-cut: by also ionizing \hi, X-rays can in fact promote \htwo\ formation \citep{gloverRadiative2003}.

While it is certainly possible to envision connections between the \CI\ emission and the Balmer break/dense gas absorption, confirming such a scenario will require more comprehensive characterization of the gas conditions in \target\ and other LRDs. 
Other far-infrared lines such as [O\,\textsc{i}]\,63\,$\mu$m or 145\,$\mu$m, which have larger critical densities than \CI, could prove a useful tracer of dense neutral gas \citep[e.g.][]{ishiiDetection2025}.
Molecular transitions in the rest-frame near-infrared, which trace warm/hot molecular gas, may be accessible with MIRI \citep[e.g.][]{costa-souzaBlowing2024}. 
Higher resolution spectra in the rest-frame UV could provide constraints on the carbon abundance \citep[e.g.][]{hsiaoFirst2024},  Ly$\beta$ pumping scenario \citep[via \oi\,$\lambda 1304$, e.g.][]{rodriguez-ardilaLine2002, matsuokaObservations2007}, and even potential detect H$_2$ fluorescence \citep[e.g.][]{johnsonTentative2025}. 
Understanding the extreme and peculiar gas conditions in LRDs will be key to characterizing their overall role in galaxy/SMBH formation.

\section{Summary}

Our observations of \CI\ and CO suggest that the LRDs do not live in particularly gas-rich host galaxies. 
This is in sharp contrast to high-$z$ quasars/obscured AGN, which have been found to reside in intensely star-forming host galaxies, routinely detected in CO, \CI, [C\,\textsc{ii}], and dust continuum \citep{venemansMolecular2017, decarliALMA2018, izumiSubaru2019, endsleyALMA2022, fujimotoDusty2022, lambridesUncovering2023}. 
Even \target, despite hosting a nearly $10^9\,M_\odot$ black hole, appears to live in a $<10^{10}\,M_\odot$ host galaxy. 
Perhaps most intriguing, the dynamical mass of \target\ is significantly lower than the stellar mass derived from SED fitting, potentially corroborating the proposed non-stellar origin of the Balmer break in LRDs.

However, given the relatively low SNR ($4.9\sigma$), it remains possible that the \CI\ detection in \target\ is spurious, or that it traces only a part of the overall host galaxy ISM, impacting the interpretation of the dynamical mass measurement. 
These results remain tentative; future deeper and/or higher resolution observations will be needed to confirm the dynamical mass measurement in \target, and reveal the host galaxy properties of other LRDs.

\section*{Acknowledgements}
	The authors wish to thank Kohei Inayoshi and Junehyoung Jeon for useful discussions that greatly improved this letter. 
	H.B.A. acknowledges the support of the UT Austin Astronomy Department and the UT Austin College of Natural Sciences through Harrington Graduate Fellowship, as well as the National Science Foundation for support through the NSF Graduate Research Fellowship Program.
	This paper makes use of the following ALMA data: ADS/JAO.ALMA\#2024.1.01085.S.
	ALMA is a partnership of ESO (representing its member states), NSF (USA) and NINS (Japan), together with NRC (Canada), NSTC and ASIAA (Taiwan), and KASI (Republic of Korea), in cooperation with the Republic of Chile. 
	The Joint ALMA Observatory is operated by ESO, AUI/NRAO and NAOJ.
	The National Radio Astronomy Observatory is a facility of the National Science Foundation operated under cooperative agreement by Associated Universities, Inc.
	
	Authors from UT Austin acknowledge that they work at an institution that sits on indigenous land. 
	The Tonkawa lived in central Texas, and the Comanche and Apache moved through this area. 
	We pay our respects to all the American Indian and Indigenous Peoples and communities who have been or have become a part of these lands and territories in Texas.

\begin{deluxetable*}{@{\extracolsep{25pt}}c@{}c@{}C@{}C@{}C@{}}[ht!]
\tabletypesize{\small}
\centering
\tablecaption{Summary of measurements.}\label{tab:measurements}
\tablehead{\colhead{Property} & \colhead{Units} & \colhead{RUBIES-BLAGN-1} & \colhead{A2744-45924} & \colhead{A2744-QSO1$^\dagger$}}
\startdata
$I_{\rm CO(7-6)}$                      & mJy\,km\,s$^{-1}$       		& <35     & <43                 & <37    \\
$L'_{\rm CO(7-6)}$                     & $10^8$~K\,km\,s$^{-1}$\,pc$^2$ 	& <3.2    & <6.9                & <11.5  \\
$I_{[{\rm C}\,\textsc{i}](2-1)}$       & mJy\,km\,s$^{-1}$       		& <41     & 36.9_{-\rm 8.4}^{+\rm 10.6} & <37    \\
$L'_{[{\rm C}\,\textsc{i}](2-1)}$      & $10^8$~K\,km\,s$^{-1}$\,pc$^2$ 	& <3.6    & 5.8^{+\rm 1.7}_{-\rm 1.4}   & <11.5  \\
$M_{{\rm H}_2,{\rm CO}}$			   & $10^{10}~M_\odot$		   		& <1.4    & <3.1                & <5.1   \\
$M_{{\rm H}_2,[{\rm C}\,\textsc{i}]}$  & $10^{10}~M_\odot$	   			& <1.1    & 2.0^{+\rm 3.1}_{-\rm 1.2}   & <3.9   \\
$M_{\rm dyn}$					       & $10^9~M_\odot$		   			& \dots   & {1.4^{+\rm 1.7}_{-\rm 0.6}}^* & \dots  \\
$S_{{\rm 370}\,\mu{\rm m}}$	           & $\mu$Jy 			   			& <29.1   & <22.5               & <2.4   \\
$M_{\rm dust}$	    		   		   & $10^7~M_\odot$	       			& <6.8    & <13.9               & <5.9   \\[0.4em]
\enddata
\tablenotetext{*}{~Dynamical mass estimated from the \CI\ line for \target, see text for details.}
\tablenotetext{\dagger}{~All measurements for A2744-QSO1 are corrected for lensing magnification.}
\tablecomments{All upper limits are 3$\sigma$. }
\end{deluxetable*}
\vspace{-0.8cm}

\newpage
\bibliographystyle{aasjournal} 
\bibliography{hollis.bib}

\end{document}